\documentclass[10pt,conference]{IEEEtran}
\IEEEoverridecommandlockouts
\usepackage{cite}
\usepackage{amsmath,amssymb,amsfonts}
\usepackage{algorithmic}
\usepackage{graphicx}
\usepackage{textcomp}
\usepackage{xcolor}
\usepackage{xspace}
\usepackage{hyperref}
\usepackage{booktabs}
\usepackage[switch]{lineno}
\def\BibTeX{{\rm B\kern-.05em{\sc i\kern-.025em b}\kern-.08em
    T\kern-.1667em\lower.7ex\hbox{E}\kern-.125emX}}

\usepackage{listings}
\usepackage{multirow}
\usepackage{float}

\lstdefinelanguage{TypeScript}{
  keywords={async, function, const, let, var, return, try, 
             catch, throw, new, if, await, export},
  keywordstyle=\bfseries,
  comment=[l]{//},
  commentstyle=\itshape,
  stringstyle={},
  morestring=[b]",
  morestring=[b]',
  morestring=[b]`
}

\lstdefinelanguage{Dart}{
  keywords={while, if, else, await, final, break, true, 
             false, null, async, return},
  keywordstyle=\bfseries,
  comment=[l]{//},
  commentstyle=\itshape,
  stringstyle={},
  morestring=[b]",
  morestring=[b]',
}

\lstdefinelanguage{JSON}{
  basicstyle=\ttfamily\small,
  string=[s]{"}{"},
  stringstyle=\color{black},
  comment=[l]{//},
  commentstyle=\color{gray}\itshape
}

\newcommand{\mcpapp}{MCPApp\xspace}
\newcommand{\mcpapps}{MCPApps\xspace}
\newcommand{\dataset}{\texttt{MCPAppDS}\xspace}
\newcommand{\taxonomy}{\texttt{MCPAppTax}\xspace}
\newcommand{\MCPImplDS}{\texttt{Toeppe-MCPImpl}\xspace}

\newcommand{\mainrq}{How do MCPApps configure and use MCP servers as software dependencies?}

\newcommand{\repo}[1]{\href{https://github.com/#1}{\texttt{#1}}}

\newcommand{\totalapps}{1,723\xspace}

\newcommand{\rqone}{\textbf{How do \mcpapps{} configure the MCP servers they depend on?}}
\newcommand{\rqtwo}{\textbf{How do \mcpapps{} manage communication with MCP servers?}}
\newcommand{\rqthree}{\textbf{What human-in-the-loop controls do \mcpapps{} place on tool invocation?}}

\begin{document}

\title{An Empirical Study of Model Context Protocol Applications}


\author{\IEEEauthorblockN{Muhammad Hamza Arshad Majeed}
\IEEEauthorblockA{\textit{New York University Abu Dhabi}\\
Abu Dhabi, United Arab Emirates \\
hamza.majeed@nyu.edu}
\and
\IEEEauthorblockN{May Mahmoud}
\IEEEauthorblockA{\textit{New York University Abu Dhabi}\\
Abu Dhabi, United Arab Emirates \\
m.mahmoud@nyu.edu}
\and
\IEEEauthorblockN{Sarah Nadi}
\IEEEauthorblockA{\textit{New York University Abu Dhabi}\\
Abu Dhabi, United Arab Emirates \\
sarah.nadi@nyu.edu}
}

\maketitle

\begin{abstract}
The Model Context Protocol (MCP) standardizes how large language model applications communicate with external tools, but leaves the application side unspecified: unlike traditional dependencies resolved through package managers, developers integrating MCP servers face no conventions for configuration, communication, or human oversight. This ecosystem is also under-researched, with existing work focused on servers rather than the applications consuming them. We conduct a large-scale study of \totalapps \mcpapps mined from GitHub. We first derive \taxonomy from a representative sample, then use an
LLM-assisted pipeline to apply it across the full dataset, characterizing
server integration across configuration, SDK use, and
human-in-the-loop mechanisms. Our results show that the ecosystem has converged on some practices but not others: most \mcpapps configure servers using files (85.2\%) and use an official SDK (81.1\%) to communicate with servers, yet no naming convention has emerged for configuration files. Human oversight diverges most, logging (90.8\%) and enable/disable controls (77.2\%) are common, but only 37.2\% gate tool execution behind a blocking approval step, leaving the LLM able to invoke any enabled tool unconditionally in most \mcpapps.
\end{abstract}


\section{Introduction}

The Model Context Protocol (MCP) is an open standard introduced by Anthropic in late 2024~\cite{anthropic2024mcp} to standardize how Aritificial Intelligence (AI) systems, specifically those powered by large language models (LLMs), connect to external tools, data sources, and services. LLMs are fundamentally constrained by their static training data and limited context window~\cite{Lewis2020}. Giving LLMs access to external tools such as databases, search engines, and web services extends their capabilities by supplying real-time context and executable functionality. MCP defines a universal interface: rather than every AI application writing custom integration code for every external tool it uses, MCP lets both sides implement a single shared protocol. Architecturally, MCP defines a client-server protocol that decouples the AI application from external resources, assigning distinct roles to the MCP host, client, and server. An \emph{MCP server} wraps an external service and exposes its functionality through a standardized schema; an \emph{MCP host} is an AI application that consumes one or more such servers through embedded \emph{MCP clients}. In this paper, we focus on MCP hosts, which we hereafter refer to as \emph{MCP-enabled AI applications}, or \emph{\mcpapp{}} for short.

In essence, MCP turns an MCP server into a \emph{dependency}: developers publish servers that expose resources and functionality, and \mcpapps{} instantiate clients to call and consume them\cite{anthropic2024mcp}. This mirrors how software systems reuse third-party libraries and web services whose behavior affects functionality, performance, security, and reliability. Most work on MCP focuses on the server side, such as server maintainability~\cite{hasan2025model} and security~\cite{Hou25}, leaving how \mcpapps{} integrate and configure servers in practice largely unexamined. This stands in contrast to traditional software dependencies, which have long been studied from the consuming application's side, e.g., how applications select, configure, and update the third-party libraries they depend on~\cite{larios2020selecting,cox2015measuring,kula2018developers}. 

By characterizing how \mcpapps{} configure, communicate with, and oversee the MCP servers they depend on, we aim to understand how the emerging MCP ecosystem is evolving and whether it raises new challenges in dependency management. We center our study on answering the following question: \emph{\mainrq{}} We decompose it into three research questions, one per integration dimension:
\begin{itemize}
  \item \textbf{RQ1:}~\rqone{} We characterize where and how \mcpapps{} declare and store the servers they depend on.
  \item \textbf{RQ2:}~\rqtwo{} We study how \mcpapps{} instantiate the MCP client and communicate with servers.
  \item \textbf{RQ3:}~\rqthree{} We examine the human oversight mechanisms \mcpapps{} implement around tool execution.
\end{itemize}

To answer these questions, we build \dataset{}, a dataset of \totalapps{} \mcpapps{}. From a representative sample, we derive the MCP Configuration Taxonomy (\taxonomy{}), which captures variations across the dimensions we study. Using an LLM-based classification pipeline, we then label the remaining \mcpapps{} according to \taxonomy{}.
Our results show that the MCP ecosystem has converged on some integration practices while remaining fragmented on others. For configuration, 85.2\% of \mcpapps{} store their server list in a file, but most use different file names. For communication, 81.1\% use an official MCP SDK, while 18.9\% implement the client layer themselves. Human-in-the-loop controls show the greatest divergence: 90.8\% log at least part of the MCP execution lifecycle and 77.2\% maintain explicit enable/disable control over servers or tools, but only 37.2\% implement a blocking approval gate before tool execution. Unlike traditional dependencies, whose declaration, resolution, and trust are mediated by package managers, MCP has no equivalent layer, leaving \mcpapps{} to manage server dependencies through application-specific conventions and pointing to a need for ecosystem-level standards as MCP matures. To the best of our knowledge, this is the first large-scale study of the MCP ecosystem that focuses on \mcpapps{} and casts a dependency lens on how they configure and use MCP servers. All scripts and data are available in our online artifact.\footnote{\url{https://figshare.com/s/1dc2ea96aca29038d99e}}

\section{Background and Motivating Examples}
\label{sec:background}


In this paper, we follow the definitions outlined by the official MCP documentation~\cite{mcp_doc}. An \textit{MCP server} is a program that can extend an LLM's capabilities and provide additional context through exposing a set of tools, or prompts that can be executed, and data sources. MCP servers can execute locally or remotely.
An \textit{MCP host} is an AI application that connects to MCP servers through MCP clients. An \textit{MCP client} is the protocol-level component of a host that enables connections to servers.
Typically, each MCP server used by a host needs a corresponding MCP client to enable connections to it.
While we have observed that the term MCP client is used loosely in practice to refer to the AI applications themselves as well \cite{Hua26}, we follow the MCP documentation that clearly separates hosts (i.e., the AI application) from the low-level clients. For clarity, we refer to the MCP host/AI application as \textit{\mcpapp}. The canonical example of an \mcpapp is Claude Desktop \cite{anthropic2024mcp}, used as the primary reference implementation by Anthropic. It can manage connections to multiple MCP servers simultaneously using MCP clients, with each client handling direct communication with a single server. 

MCP standardizes the protocol by which servers communicate with clients within \textit{\mcpapps}, including the JSON-RPC 2.0 message format, handshake and capability negotiation, as well as primitives exposed by the server (tools, prompts, and resources). MCP also standardizes the transport mechanisms (Stdio and Streamable HTTP) utilized. Accordingly, \textit{MCP SDKs} are language-specific libraries that implement the full MCP specification, allowing developers to easily implement MCP servers and clients.
Official MCP SDKs exist for TypeScript~\cite{typescriptsdk}, Python~\cite{pythonsdk}, Java~\cite{javasdk}, Kotlin~\cite{kotlinsdk}, C\#~\cite{csharpsdk}, Swift~\cite{swiftsdk} and Rust~\cite{rustsdk}.

While MCP standardizes client-server communication, it leaves \mcpapp-side architectural and operational boundaries unspecified. For example, there is no single enforced way to declare or specify the servers available to an \mcpapp. Additionally, while MCP SDKs exist, developers are not forced to use them and can implement the communication and transport mechanisms themselves (as long as they adhere to the MCP specification). Finally, the MCP standard does not specify any approval or human-in-the-loop (HITL) mechanisms that should be performed before a tool from an MCP server is run.

To illustrate these variances, we provide real motivating examples found from our analysis. Listing~\ref{lst:configexample} shows how the \textit{\mcpapp} \repo{ChatGPTNextWeb/NextChat}~\cite{nextchat},reads from a JSON configuration file (\texttt{mcp\_config.json}) to discover the list of servers to connect to, alongside their configuration parameters such as the command, arguments, or transport settings. 
Listing~\ref{lst:clientexample} shows how the application then imports the official MCP SDK client and transports, i.e., the \texttt{Client} and \texttt{StdioClientTransport} classes. Crucially, during the execution of tools on Lines 10 -14, the LLM request is routed directly to the tool call, with no human intervention to approve or deny the process; the tool call is only logged.

\begin{figure}[t!]
\begin{lstlisting}[language=JSON, caption={MCP server configuration file in \\ \repo{ChatGPTNextWeb/NextChat}.},label={lst:configexample}]
{
  "mcpServers": {
    "filesystem": {
      "command": "npx",
      "args": ["-y", "@modelcontextprotocol/server-filesystem",
               "/Users/alice/projects"],
      "env": {}
    }
  }
}
\end{lstlisting}
\end{figure}

\begin{figure}[t!]
\begin{lstlisting}[language=TypeScript,  caption={MCP client instantiation and ungated tool call in \\ \repo{ChatGPTNextWeb/NextChat} (adapted to shorten)}, label={lst:clientexample}, 
  basicstyle=\ttfamily\footnotesize,
  columns=fullflexible,
  keepspaces=true,
  mathescape=false,
  literate={`}{{\`{}}}1 {$}{{\$}}1]
// client.ts - SDK-based client construction
import { Client } from "@modelcontextprotocol/sdk/client/index.js";
import { StdioClientTransport } from "@modelcontextprotocol/sdk/client/stdio.js";

const transport = new StdioClientTransport({ command, args, env });
const client = new Client({ name: "nextchat-mcp-client" }, {});
await client.connect(transport);

// actions.ts - ungated tool execution
export async function executeMcpAction(clientId, request) {
  const client = clientsMap.get(clientId);
  logger.info(`Executing request for [${clientId}]`);
  return await executeRequest(client.client, request); // no approval gate
}
\end{lstlisting}

\end{figure}

On the other hand, \repo{daodao97/ChatMCP}~\cite{chatmcp} stores its server list in two locations depending on the platform. On desktop and mobile, it reads and writes a JSON file called \texttt{mcp\_server.json} in the application data directory, with default values bundled as an asset (\texttt{assets/mcp\_server.json}), while on the web, it stores the list under the key \texttt{mcp\_servers\_json} in SharedPreferences (browser local storage). As shown in Listing~\ref{lst:selfimpl}, the client is implemented entirely by the application itself rather than using the official MCP SDK: On Lines 1-7, \texttt{StdioClient} constructs raw JSON-RPC 2.0 \texttt{tools/call} messages manually and writes them directly to the server process. ChatMCP also provides both per-server and per-tool enabling toggles. Furthermore, as shown in Listing~\ref{lst:selfimpl} Lines 12-17, ChatMCP implements a dedicated human-in-the-loop flow in \texttt{\_showFunctionApprovalDialog()}, where it displays a non-dismissible \texttt{AlertDialog} before tool execution, which the user must explicitly allow or cancel.

\begin{figure}[t!]
\begin{lstlisting}[language=Dart,
  basicstyle=\ttfamily\footnotesize,
  columns=fullflexible, keepspaces=true,
  caption={Self-managed client and approval gate in ChatMCP.},label={lst:selfimpl}]
// stdio_client.dart - manual JSON-RPC message construction (no SDK)
Future<JSONRPCMessage> sendToolCall({required String name,
    required Map<String, dynamic> arguments}) async {
  final message = JSONRPCMessage(
    method: 'tools/call',
    params: {'name': name, 'arguments': arguments},
  );
  return sendMessage(message); // writes raw JSON to server process
}

// chat_page.dart - blocking approval gate before execution
final approved = await _showFunctionApprovalDialog(event);
if (approved) {
  await _sendToolCallAndProcessResponse(event.name, event.arguments);
} else {
  _runFunctionEvents.clear(); // user cancelled
}
\end{lstlisting}
\end{figure}


The examples above show that developers vary in how they manage MCP server integration in practice, a marked contrast to the integration of ``traditional'' third-party packages. Each programming language ecosystem typically has package managers that enforce how dependencies are declared (e.g., \texttt{pom.xml} for Maven with Java, \texttt{package.json} for JavaScript), yet no comparable standard governs how an \mcpapp{} declares the MCP servers it depends on. Beyond dependency declaration, MCP introduces unique architectural considerations that have no equivalent in traditional dependencies. The first is communication: whereas a library can only be invoked directly through API calls in the programming language, an MCP server is reached over a wire protocol. The MCP SDK provides high-level APIs for this, but using it is optional. Since MCP is a wire protocol, compliance follows only from adhering to the specification (e.g., the JSON-RPC 2.0 message schemas), allowing an \mcpapp{} to forgo the SDK and implement its own client. The second is invocation: while explicit calls in code determine which (and when) of a library's APIs are used, an \mcpapp{} has no such control by default over when a server's tools get executed. Instead, the LLM decides which tools to call when. Governing these calls requires adding human-in-the-loop mechanisms. In this paper, we study how these differences manifest in practice across MCP-enabled applications.



\section{Dataset Construction} \label{sec:dataset}

To answer our research questions, we need to identify a large collection of \mcpapps to analyze. Toeppe et al.~\cite{MSRdataset} recently released a dataset of GitHub repositories with MCP-relevant implementations, including both client and server repositories (referred to hereafter as \MCPImplDS).
Specifically, this dataset contains 2,297 GitHub repositories, of which the authors classify 711 as \textit{Server}, 215 as \textit{Client}, and 1,240 as \textit{Both}. Given our terminology defined in Section~\ref{sec:background}, the clients' category corresponds to the \mcpapps we focus on.
Thus, we are specifically interested in the repositories they categorize as ``client'' or ``both'', which amounts to 1,455 repositories.
However, upon inspecting the repositories listed in \MCPImplDS, we notice several limitations. 
First, the dataset contains only repositories created or updated between January 2024 and October 2025.
Additionally, upon further inspection of the repositories marked as clients or both, we notice several mismatches in the categorization. For example, \repo{google-gemini/gemini-cli}~\cite{geminicli} is classified in the dataset as ``both'' with high confidence. However, based on its documentation, \textit{``Gemini CLI brings the power of Gemini models directly into your terminal''} with the \textit{``Setup an MCP server''} section further providing guidance on how to \textit{``connect Gemini CLI to [\ldots] external databases and services.''}. This indicates that Gemini CLI should be categorized as client only (or \textit{MCPApp} in our terminology), not both. 
Another example is the repository \repo{IBM/mcp}~\cite{ibmmcp}, which is categorized as ``both''. However, this repo only documents a list of IBM MCP Servers, a list of contributors, and a link to a Discord group. For information relevant to \mcpapps, the repository only has this line: \textit{``We recommend using Langflow or your IDE of choice as MCP client.''} Accordingly, this is a documentation repository that does not actually contain the implementation of an \mcpapp. Another example is of \repo{modelcontextprotocol/python-sdk}~\cite{pythonsdk}, which is again classified as ``both'' in the dataset while it is neither an \mcpapp nor a server, but is instead an official SDK that used for building \textit{MCPApps} or MCP Servers. 

Based on these observations, we define a new mining pipeline to precisely identify \mcpapps.
Such a pipeline follows the same idea of a two-step process used in \MCPImplDS: searching GitHub for candidate repositories, followed by further filtering and categorization.
However, we are mainly interested in finding as many \mcpapps as possible (rather than any MCP-related repository).
We first conduct a detailed manual investigation to identify (1) search keywords we can use on GitHub and (2) filtering criteria that can be used to identify \mcpapps.
We then design an automated search and filtering pipeline to construct our final data set.
As a result of our manual investigation, we also construct a ground-truth labeled dataset to evaluate our automated pipeline. 

In the next subsections, we first describe our manual investigation that led to the final search queries that we used to mine GitHub for candidate repos, and then we detail our automated approach for classifying repositories as \mcpapps.

\subsection{Manual Investigation and Identifying Candidate \mcpapp Repositories from GitHub}

\paragraph{Deriving search queries}

\MCPImplDS used 5 GitHub search queries to identify MCP-relevant repos (\textit{``MCP'', ``mcp server'', ``Model Context Protocol'', ``Claude Desktop MCP'', ``mcp client''}). They searched for these keywords in the repository name, description, and README \cite{MSRdataset}.
Since our objective is to target \mcpapp implementations, we need search terms that are targeted specifically towards MCP hosts and clients, rather than any MCP-related repository.
Accordingly, we inspect a subset of repositories from \MCPImplDS that we manually confirmed as \mcpapp.
We look into recurring keywords found in repositories that would be considered \mcpapp. We also identify some additional open-source repositories in registries that list \mcpapps, including Glama, MCP.so, PulseMCP, and MCPservers.org\footnote{\url{https://glama.ai/mcp/clients}; \url{https://mcp.so/clients}; \url{https://www.pulsemcp.com/clients}; \url{https://mcpservers.org/clients}}, as well as Awesome MCP Clients~\cite{awesomemcpclients} and examples on the official MCP documentation. Some of the repositories we inspected include \repo{nanbingxyz/5ire}~\cite{fiveire} and \repo{aaif-goose/goose}~\cite{goose}. We inspect each of these repositories' README and source code to identify any recurrent patterns that we can use to search GitHub for potential \mcpapps repositories.

\begin{table}[t]
\centering
\caption{Examples of top-yielding GitHub search queries, drawn from the 17 repo-search and 29 code-search queries.}
\label{tab:github-search}
\begin{tabular}{p{0.15\columnwidth}p{0.75\columnwidth}}
\toprule
\textbf{Search type} & \textbf{Example queries} \\
\midrule
repo search & \texttt{Model Context Protocol in:name,description,readme}, \texttt{MCP client in:name,description,readme}, \texttt{mcp-client in:name,description,readme}, \texttt{mcp-go in:name,description,readme}, \texttt{topic:mcp} \\
\addlinespace
code search & \texttt{client/streamableHttp.js language:TypeScript}, \texttt{SSEClientTransport language:TypeScript}, \texttt{mark3labs/mcp-go filename:go.mod}, \texttt{client/stdio.js language:TypeScript}, \texttt{from mcp.client.streamable\_http language:Python} \\
\bottomrule
\end{tabular}
\end{table}

Through this manual analysis, we identify both keywords that appear in the repository documentation or descriptions and specific API calls that typically occur in \mcpapps.
Specifically, we identify 17 keywords that can appear in a repo's topics, name, description, or README file. For example,  \textit{``Model Context Protocol''}, \textit{``MCP Host''}, \textit{``MCP Application''} and \textit{``MCP Client''}.
We also identify 29 APIs or dependencies that can appear in the codebase, such as \textit{``@modelcontextprotocol/sdk''}, or transport patterns, e.g., \textit{``StdioClientTransport''}.
Table~\ref{tab:github-search} shows examples of the top-yielding queries among the 46 we used; the full list is included in our online artifact.

\paragraph{Inspect Candidates and Build Ground Truth Dataset}
Even though our search terms are geared towards \mcpapps, it is natural that the search may return MCP servers or even non-MCP relevant repositories. Accordingly, we need to define a filtering strategy to identify \mcpapps from the set of candidate repositories. Executing the GitHub search queries identified in the previous step returns 21,355 repos. Following \MCPImplDS, we set the minimum repository update date to January 2024, allowing us to retrieve repos with at least one update between January 2024 and until the search execution date (latest June 11, 2026). We run all 46 queries, merge the results, and de-duplicate repositories returned by more than one query. To filter out small-scale toy projects, we also discard repositories with fewer than 10 stars or a size smaller than 10\,KB. Through this sequence of steps, we obtain 6,994 candidate repositories.

We select a statistically representative random sample of 68 candidate repositories (90\% confidence level and 10\% margin of error) for manual analysis. Through our manual analysis, we first want to confirm which of these are actually \mcpapps, providing us with ground truth we can use to evaluate any automatic categorization. Secondly, we keep track of the reasons (or \textit{evidence}) we find in the repositories that led us to identify them as \mcpapps. This evidence can serve as filtering criteria for an automated pipeline. Two of the authors independently inspect each sampled repository to determine whether it is an \mcpapp and record the evidence they use to reach their decision. If it is not an \mcpapp, they mark the repository type it best matches. Overall, we identify the following \texttt{RepoType} categories for the returned candidate repos: \mcpapp, MCP Server, SDK, and Documentation.
There were even some Non-MCP Relevant repositories returned, such as \repo{spmallick/learnopencv}~\cite{learnopencv}, which is a computer vision tutorial.
Out of the 68 examined repositories, we found only 21 to be \mcpapps. The labeled set of 68 repositories serves as our ground truth for later automated categorization.

\subsection{Automatic Classification of \mcpapps{}}
Our analysis showed that no single keyword, dependency, or manifest can cleanly and deterministically separate these categories using a rule-based filtering method. The same MCP import or transport class can appear across servers, SDKs, and Apps, so assigning a category requires reading how MCP is actually used. Since manually reviewing the code across the entire candidate dataset is infeasible, we define an LLM-based approach to categorize repositories into the \textit{RepoTypes} we identified above. For each repository, we clone it and scan all source and configuration files for MCP-related signatures, ranking files by hit count. We then assemble the README and the top-ranked files into a single structured prompt that asks the model to categorize the repository into one of the five \texttt{RepoType} categories, or return \textit{Unsure} if the evidence is insufficient. If the model returns \textit{Unsure}, we perform a second pass with additional evidence from the ranked list. A second \textit{Unsure} is accepted as final. We use \texttt{gpt-5} for this categorization, with our full prompt available in our artifact. Before running our LLM-based classification pipeline on the full dataset, we first evaluate it on our 68 manually labeled repositories. The classifier achieves an overall accuracy of 95.6\% (65/68). For MCPApp identification in particular, the precision is 1.000, recall 0.905, and the F1 score is 0.950.


\subsection{Final Data Set}
Having validated our pipeline's performance on a ground-truth set, we then apply it to the full set of mined repositories. At the time of running the classification script, four of the repository URLs we collected were no longer available and could not be cloned. Accordingly, the final input list to the LLM contained 6,990 repositories. 
Apart from 20 repositories that required a second iteration (i.e., first iteration returned Unsure), the LLM classified all repositories in the first attempt.

Overall, the LLM categorizes 1,727 repos as MCPApps, 2,732 as MCP-Servers, 522 as SDKs, 823 as documentation, and 1,181 as Non-MCP-relevant. There were also 5 remaining Unsures, which we manually resolve. None of these turned out to be MCPApps. As a final additional validation, we randomly sample an additional 30 repositories from the 1,727 repositories labeled as \mcpapps. We manually confirm that 29 of these 30 are indeed \mcpapps, while the remaining one is a documentation repository that contains a small MCP app demo implementation snippet. This gives us high confidence that our classification pipeline is robust and that we can proceed with using it. We discard the documentation repository, leaving us with 1,726 \mcpapps. At the time of our final analysis, 3 of these repositories were deleted or set to private. We exclude these three, leaving us with a final data set of 1,723 \mcpapps referred to as \dataset.

\section{Methods: Analyzing \mcpapps}\label{sec:taxonomy}


 We design a multi-stage research pipeline to identify and categorize the integration patterns of MCP servers in \mcpapps.
Our approach follows a derive-then-scale structure: we first manually analyze 50 randomly sampled \mcpapps{} to derive \taxonomy{}, then develop an automated LLM-based pipeline that applies \taxonomy{} across the full dataset.

\subsection{Manual Analysis and Taxonomy Derivation}
We first manually inspect 50 randomly sampled repositories from our \dataset. 
We review the source code, configuration files, and documentation of each repository to answer each of our three research questions. During the review, we note the evidence we find so we can use it later for our large-scale LLM-based analysis. We iteratively combine observations into named codes/categories. As we examined more repositories, the classification properties began to repeat, and no new categories emerged. 
Specifically, we find that the taxonomy reached empirical saturation at about 30 repositories.
Based on this analysis, we derive the MCP integration taxonomy, \taxonomy. 

Two authors independently labeled 29 of the 50 repositories across all
five \taxonomy{} dimensions, with per-dimension raw agreement between
75.9\% and 93.1\%. Because per-dimension Krippendorff's $\alpha$ can be misleadingly low when
label distributions are skewed, we report a pooled
$\alpha$ over all $29 \times 5 = 145$ (repository, dimension) items: 0.82,
indicating substantial agreement. We resolved all disagreements, which were mostly
boundary cases, through discussion; one author then labeled the remaining
20 repositories using the refined criteria.


\subsubsection{Dimension 1: Configuration}
We find that there are mainly two ways in which MCP servers are declared and configured:

\textbf{Configuration Files:} Server definitions are stored in an external, non-executable file such as JSON, YAML, or TOML. The file typically follows a declarative schema listing server names alongside their transport type, command, arguments, and environment variables. A sample configuration file can be seen in Listing~1. For example, \repo{johnrobinsn/askit}~\cite{askit} stores its server list in \texttt{mcp\_config.json} while \repo{mario-andreschak/FLUJO}~\cite{flujo} stores it in \texttt{mcp\_servers.json}. 

\textbf{Database:} Here, server definitions are stored in a relational or embedded database and managed through explicit Object-Relational Mapping (ORM) or query operations. For example, \repo{nanbingxyz/5ire}~\cite{fiveire}
stores server records in a PGlite table via Drizzle ORM. We also observe server definitions declared in client-side stores such as \texttt{LocalStorage} or \texttt{IndexedDB} in web-native and desktop hybrid applications (e.g., Electron, Chrome extensions). We consider these as a form of database as well. For example, \repo{NitroRCr/AIaW}~\cite{aiaw} stores its server list using Dexie.js with \texttt{installedPluginsV2} as the keyed store.

\subsubsection{Dimension 2: Client Instantiation and Server Communication} Recall that the \mcpapp communicates with an MCP server through an initiated MCP client within the app. We find that there are mainly two ways by which this process is managed:

\textbf{The MCP SDK:} Here, the \mcpapp creates the client and manages server communication all through the MCP SDKs for each programming language. For example, \repo{johnrobinsn/askit}~\cite{askit} uses the official MCP SDK methods for Python, specifically \texttt{stdio\_client(..)}, \texttt{sse\_client(..)}, \texttt{streamablehttp\_client(..)}.

\textbf{Self-Managed:} In this setup, the \mcpapp does not actually use the official 
MCP SDK to communicate with the server. Instead, it writes its own client-server 
communication code.  For example, \repo{daodao97/ChatMCP}~\cite{chatmcp} (discussed in
Section~\ref{sec:background} and Listing~3) defines a custom \texttt{McpClient} interface with four concrete implementations selected at runtime based on the server type: 
\texttt{StdioClient}, \texttt{SSEClient}, \texttt{StreamableClient}, and 
\texttt{InMemoryClient}. Each implementation handles its own transport-level communication, JSON-RPC message construction, and session lifecycle independently.

\subsubsection{Dimension 3: Human-in-the-loop (HITL)}
In this dimension, we describe whether an \mcpapp includes any mechanisms through which human oversight is incorporated into the tool invocation workflow. We identify three mechanisms that may be present in any combination: \textit{approval required}, \textit{allowed list}, and \textit{logging}.


\textbf{Approval Required:} The \mcpapp intercepts an MCP tool call before execution and explicitly prompts the user to approve or deny it. The tool does not execute until the user responds. For example, before each tool call, \repo{nanbingxyz/5ire}~\cite{fiveire} asks the user for confirmation where they can allow the tool to execute always, never, or only once (i.e., it will ask again the next time the tool is executed). Similarly, \repo{autohandai/code-cli}~\cite{codecli} implements a \texttt{PermissionManager} with \texttt{mode:'interactive'} as its default, halting on each tool call. Approval Required applies only when the source code clearly show such a blocking user approval step gates tool execution.

\textbf{Allowed List:} The \mcpapp maintains a user-controlled list or flag that 
determines which servers or tools are active, checked before or during invocation.
Unlike Approval Required, which intercepts each tool call in real time, an allowed
list is configured in advance, so once a server or tool is enabled, it executes
without further prompting. For example, \repo{AstrBotDevs/AstrBot}~\cite{astrbot} implements  an \texttt{active} flag on each server entry controlled via a dashboard toggle  (\texttt{enable\_mcp\_server}). A repository is considered to use Allowed List only when there is an explicit run-time check that can reject or skip a server or tool based on user-controlled configuration.

\textbf{Logging:} The \mcpapp logs MCP tool activity, traces, or monitoring dashboards under its default runtime configuration. In this case, the code has logging infrastructure 
active at INFO level during the MCP connection lifecycle, tool calls, or error handling. 
This as opposed to DEBUG level logging, which is only used by the app developers themselves.
For example, \repo{daodao97/ChatMCP}~\cite{chatmcp} uses the Dart \texttt{Logger} package extensively 
throughout its MCP connection lifecycle and tool invocation paths. 

\subsubsection{Use Case Domain and Domain Specificity}
In addition to the three integration dimensions above, we also examine how \mcpapps vary in their primary purpose and target
domain. We therefore collect two supplementary labels for each repository.
\textbf{Use Case Domain} captures the application's primary role: an
\textit{AI Assistant} is a standalone tool that end users run to accomplish
tasks using AI; an \textit{MCP Support Tool} is used by developers to build,
test, evaluate, or debug an AI or MCP system; and \textit{Infrastructure}
refers to components that manage or proxy MCP servers themselves rather than
serving end users directly.
\textbf{Domain Specificity} captures whether the application targets a
particular domain (e.g., \textit{finance}, \textit{healthcare},
\textit{software development}) or is \textit{general-purpose}. A
general-purpose host is not built around any one domain, it's usually a chatbot that users can point at whatever they need by connecting the MCP servers of their choice.

\subsection{LLM-based Classification Pipeline}

For feasibility of categorizing all \mcpapps according to \taxonomy dimensions, we create an LLM-based classification pipeline. The central challenge is \emph{evidence retrieval}: a repository may contain thousands of files, but the classification along any dimension can typically be decided by a small handful of files. Supplying the entire repository to the model is both expensive and counterproductive, as the few decisive files are drowned out by irrelevant context. Our pipeline therefore first retrieves a compact set of evidence files per taxonomy dimension, and then issues 2 focused LLM queries, one for configuration, communication and the use case domain, and another for human-in-the-loop.

\paragraph{Evidence Retrieval}
Using the cloned repository, we analyze all files while skipping generated files or dependency directories 
such as \texttt{node\_modules}, \texttt{dist}, \texttt{.venv}, and \texttt{target}.
We maintain two independent keyword sets that determine which files are most relevant 
for each query. The \textit{MCP keyword set} combines configuration-related terms 
(e.g., \texttt{mcpServers}, \texttt{mcp\_server.json}, \texttt{yaml.load}, 
\texttt{sqlite}, \texttt{localstorage}, \texttt{drizzle}, \texttt{indexeddb}) with 
communication-related terms (e.g., \texttt{StdioClientTransport}, \texttt{client.connect}, 
\texttt{callTool}, \texttt{JSONRPCMessage}). The 
\textit{HITL keyword set} covers oversight-related terms such as \texttt{approval}, 
\texttt{permission}, \texttt{allowed\_list}, and \texttt{logging}. Each file is 
scored by the total number of keyword hits in its content, producing two ranked lists, one for each of the evidence buckets.

\paragraph{File Selection}
For the MCP query (configuration and communication), we select up to 20 files from 
the ranked list using a three-tier priority: (1) files whose path or filename contains  \texttt{mcp}, (2) files whose content contains \texttt{mcp}, and (3) remaining files in keyword hit rank order. This reflects our observation that MCP-specific integration code is most often located in MCP-named files. For the HITL query, we use a two-tier  priority: (1) files whose content contains \texttt{mcp} (since HITL logic frequently  lives in general-purpose UI or controller code whose path does not mention MCP), and  (2) remaining files in keyword hit rank order. 
For each query, we include the selected files (truncated to 20,000 characters each) and the README file (truncated to 4,000 characters).

\paragraph{LLM Query 1: Configuration and Communication}
The first query presents the selected MCP evidence files together with the
README and asks the model to assign the two supplementary labels, use-case domain and domain specificity, as well as the two \taxonomy dimensions
covered by this query: configuration type, including the concrete file or
database name, and communication source, including the concrete client or
session class names used. The prompt instructs the model to classify based on code evidence for each label. If the evidence is insufficient to classify any dimension confidently, the 
model is instructed to return \textit{Unsure} rather than guess.

\paragraph{LLM Query 2: Human-in-the-loop}
The second query presents the HITL evidence files and asks the model to classify three  independent binary controls. We ask the model to classify \textit{Logging} as \textit{Yes} if the codebase has logging infrastructure active at INFO level for the MCP connection lifecycle, tool registration, or error handling. We ask it to classify \textit{Allowed List} as \textit{Yes} if there is an explicit runtime check that can reject or skip a server or tool based on user-controlled configuration. Finally, we ask it to classify \textit{Approval Required} as \textit{Yes} if the files clearly show a blocking user-approval step 
that gates tool execution. Note that the LLM categorizes each separately, allowing a repository to exhibit any combination of the three.

\paragraph{Model Selection}
We evaluated two factors when selecting the model for the classifier: whether reducing the number of files from 20 to 10 affects 
classification quality, and how \texttt{gpt-5} compares to \texttt{gpt-5-mini}. 
Running four experiments (one per combination) on the same 10 
ground-truth repositories, we found that the main taxonomy labels were largely 
insensitive to model choice, with both models agreeing on 95\% of categorical 
classifications. Reducing context from 20 to 10 files, however, reduced recall as 
decisive files were dropped from the evidence set. We therefore adopt \texttt{gpt-5-mini} with 20 files per query, substantially 
reducing cost relative to \texttt{gpt-5}.

\paragraph{Unsures}
It is possible in some cases that the evidence set sent to the LLM does not contain files that can conclusively be used for classification. For this reason, and to reduce the risk of the LLM confidently misclassifying repositories, we add an option to return `Unsure' for any of the three dimensions. We try to manually resolve as many `Unsure' labels as possible.

\subsection{Evaluation}

We evaluate the LLM-based classification pipeline against the manually annotated
ground-truth set of 50 \mcpapps from our formative study.
For each \taxonomy dimension, we compute accuracy over labelled cases (excluding entries where predicted label is \textit{Unsure}). The pipeline achieves an overall accuracy of 98.3\% across the five dimensions,
with individual axes ranging from 95.8\% (Configuration) to
100.0\% (Communication Source, Logging 
the full dataset.

We apply our LLM-based pipeline to all 1,723 \mcpapps in \dataset. After running the pipeline, 529 repositories had at least one \textit{Unsure} field. From these, we are able to resolve 471 of the repositories. For each taxonomy dimension, we report results over the repositories for which
a label was resolved; the count of resolved repositories varies slightly by
dimension and is reported in each table.
The remaining \textit{Unsure} values account for no more than 2.1\% of any
single field, and these correspond to repositories whose source code is inaccessible
(binary releases, VS Code extensions, docs-only repositories) or whose
configuration falls outside \taxonomy (CLI-supplied or hardcoded server
definitions, discussed in Section~\ref{sec:discussion}).

To further strengthen the reliability of the LLM results, we manually validated the classifications for 30  randomly sampled repositories across \taxonomy. Excluding cases where the LLM returned \textit{Unsure}, the pipeline achieves an overall accuracy of 96.5\%, with per-dimension accuracy ranging from 93.1\% to 100\%. We resolved the mismatches in the \dataset.

\section{Results} \label{sec:results}

\subsection{Dataset Overview: Use Case Domain}
 
Before examining the three taxonomy dimensions, we describe the composition of
\dataset by use case domain. We find that 68.7\% of \mcpapps are \textit{AI Assistants}, i.e., standalone applications that end users run to accomplish tasks with AI assistance. \textit{MCP Support Tools} represent 21.6\% of the \mcpapps where developers use them to build, test, or explore AI and MCP systems.
The remaining 9.7\% are \textit{Infrastructure} components
that manage or proxy MCP servers rather than serving end users directly. In terms of the domain across all \mcpapps{}, 59.5\% are general-purpose, while
the remainder target specific domains, most commonly software development (15.8\%), security (3.2\%), research (1.9\%), and finance (1.5\%).

\subsection{RQ1: How are MCP Servers Configured in \mcpapps?}
 
Table~\ref{tab:config-results} summarizes the configuration dimension across the
1,687 repositories with a resolved label.
 
\begin{table}[t!]
\centering
\caption{Distribution of Configuration types across \dataset ($n = 1{,}687$ labelled).}
\label{tab:config-results}
\begin{tabular}{lrr}
\toprule
\textbf{Configuration} & \textbf{Count} & \textbf{\%} \\
\midrule
File only        & 1,290 & 76.5\% \\
File + Database  &   148 &  8.8\% \\
Database only    &   249 & 14.8\% \\
\bottomrule
\end{tabular}
\end{table}

We find that 85.2\% of \mcpapps use a file to store their MCP server list,
either exclusively (76.5\%) or alongside a database (8.8\%).
File-based configurations are declarative JSON, YAML, or TOML documents that list
server names with their transport type, command, arguments, and environment variables. However, there is no dominant naming convention: while 69.1\% of
file-configured repositories include \texttt{mcp} in their
configuration file name, they split across competing variants
such as \texttt{mcp.json} (30.7\%) and \texttt{mcp\_servers.json}
(12.8\%). This contrasts with traditional dependency management, where file names are standardized by convention (e.g., \texttt{package.json}, \texttt{pom.xml}). Accordingly, \mcpapps{} converge on configuring MCP servers mainly through files, but not on any shared naming convention for those files.

We also find that 23.5\% of \mcpapps use a database, either as the sole mechanism or combined with a file. The database layer itself varies between SQL-backed stores and browser-native persistence (localStorage and IndexedDB, typically in Electron or web-based clients). Repositories that combine both file and database (8.8\%) typically use
files for per-project server definitions while storing user-managed or dynamically discovered servers in a database.

 

\begin{center}
\fbox{\parbox{0.95\columnwidth}{%
\textbf{RQ1 Summary:} Files are the dominant configuration mechanism (85.2\%),
but there is a lack of naming convention. Database storage is also used (23.5\%) but varies between SQL stores and browser-native persistence.
}}
\end{center}

\subsection{RQ2: How is Communication with MCP Servers Managed?}

Table~\ref{tab:comm-results} summarizes the communication sources across the
1,710 repositories with a resolved label.

\begin{table}[t!]
\centering
\caption{Distribution of Communication Source across \dataset ($n = 1{,}710$ labelled)}
\label{tab:comm-results}
\begin{tabular}{lrr}
\toprule
\textbf{Communication Source} & \textbf{Count} & \textbf{\%} \\
\midrule
SDK           & 1,386 & 81.1\% \\
Self-managed  &   324 & 18.9\% \\
\bottomrule
\end{tabular}
\end{table}
 
Among these \mcpapps,
81.1\% communicate with MCP servers through an
official MCP SDK. The remaining 18.9\% of \mcpapps implement the client layer entirely themselves, managing JSON-RPC 2.0 message construction, transport initialization, and session lifecycle without relying on any official SDK.
These repositories define custom client types (e.g., \texttt{StdioClient}, \texttt{SSEClient}, \texttt{McpClient}) that interact with MCP servers directly at the message level. The implementation is therefore valid as long as it adheres to the JSON-RPC 2.0 message format and transport conventions. 
 
\begin{center}
\fbox{\parbox{0.95\columnwidth}{%
\textbf{RQ2 Summary:} 81.1\% of \mcpapps use an official MCP SDK,
but 18.9\% implement the client layer themselves.
}}
\end{center}
 
\subsection{RQ3: What Human-in-the-Loop Controls Are Present?}

 Table~\ref{tab:hitl-results} reports the per-mechanism breakdown of observed HITL controls.
Note that the three HITL categories are independent and can appear in any combination.
 
We find that logging is nearly universal where
90.8\% of \mcpapps send logs at INFO level or above covering during at least part of the
MCP lifecycle, whether connection or tool invocation. Recall that logging does not prevent or gate any tool execution, but it provides an observability channel, either through user-facing dashboards and trace views or through application logs.
 
We also find that allowed lists are common, but approval gates are less common.
Specifically, 77.2\% of \mcpapps maintain some form of explicit enable/disable control over
servers or tools. This control is configured in advance: at each tool
call, the application automatically checks whether the requested server or
tool is enabled, without involving the user. In contrast, only 37.2\%
implement a blocking approval gate, where each tool call is suspended until
the user explicitly allows or denies it. Conversely, in 62.8\% of \mcpapps, no approval is required before a tool
executes: once a tool is enabled, the LLM's request alone triggers execution. This gap suggests that the majority of developers who implement HITL controls prefer
to set these configurations before the session, rather than through real-time interruption of the execution loop. 

\begin{table}[t!]
\centering
\caption{Distribution of HITL mechanisms across \dataset.
Each mechanism is computed over its own resolved subset ($n$ = Yes + No, excluding Unsure).}
\label{tab:hitl-results}
\begin{tabular}{lrr}
\toprule
\textbf{HITL Mechanism} & \textbf{Yes} & \textbf{No} \\
\midrule
Logging ($n=1{,}714$)           & 1,557 (90.8\%) & 157 (9.2\%) \\
Allowed List ($n=1{,}717$)      & 1,325 (77.2\%) & 392 (22.8\%) \\
Approval Required ($n=1{,}714$) &   638 (37.2\%) & 1,076 (62.8\%) \\
\bottomrule
\end{tabular}
\end{table}

We also explore the combinations of HITL mechanisms, focusing on the 1,703 repositories with resolved labels across the the three mechanisms.
From these \mcpapps, we find that
32.3\% implement all three controls simultaneously.
The most common pattern is logging combined with an allowed list but
without an approval gate (40.0\%), reflecting passive observability plus
pre-session access control with no additional approval required during tool calls. 20.0\% of \mcpapps have no active gate of any kind: neither
an allowed list nor an approval step. Within this group, 16.4\% rely on
logging alone, and 3.5\% implement no HITL mechanism at all.
 
\begin{center}
\fbox{\parbox{0.95\columnwidth}{%
\textbf{RQ3 Summary:} Logging is near-universal (90.8\%) and allowed
lists are common (77.2\%), but only 37.2\% of \mcpapps implement a blocking approval
gate. The most common pattern, logging plus an allowed list but no
approval gate (40.0\%), provides pre-session control with no real-time
intervention.
}}
\end{center}
 
\section{Discussion}
\label{sec:discussion}

\subsection{MCP Servers are not fully standardized dependencies}
 
Our central motivation is to understand how MCP servers compare to traditional software dependencies from a dependency management perspective. Our results reveal a picture of \emph{partial standardization}: the MCP ecosystem
has converged on some practices but remains fragmented on others, with meaningful dependency management and security implications along each divergence.
 \paragraph{Where standardization has emerged}
Two dimensions show clear convergence. First, files are the dominant configuration mechanism: 85.2\% of \mcpapps store their server list in a JSON, YAML, or TOML file, echoing how traditional package
managers use declarative manifests. Second, the official MCP SDKs have achieved strong adoption, with 81.1\% of \mcpapps using one to manage client-server communication. Both patterns suggest that developers have gravitated towards standardization similar to traditional software ecosystems.
\paragraph{Where standardization is absent}
Despite converging on \emph{files}, developers have not converged on a
\emph{file name}. In traditional ecosystems, the manifest file name is fixed by the package manager (\texttt{package.json}, \texttt{pom.xml}, \texttt{requirements.txt}), which enables
automated analysis tools, vulnerability scanners, or CI pipelines to reliably locate and parse dependency declarations. With MCP, no equivalent convention exists: although \texttt{mcp.json}
is the most common configuration file name, it appears in only 30.7\%
of file-configured repositories. Tooling that wants to audit which MCP servers an application depends on cannot assume any specific file name, making automated dependency discovery significantly
harder. Similarly, 81.1\% use an official SDK while 18.9\% implement the MCP client layer themselves. Traditional in-process libraries surface no such layer: developers call an API directly, with no protocol to implement, so the choice does not arise. Because MCP is a wire protocol, a self-managed client is valid, but it also means that a fraction of \mcpapps{} carry custom JSON-RPC 2.0 implementations that may deviate from the specification.
 \paragraph{Security implications of missing oversight standards}
With MCP, the decision of which tools get called is delegated to the LLM, and the absence of a standardized oversight mechanism leaves the \mcpapp developers to decide independently how much control to impose. Our results show that these choices vary, spanning the full range from blocking approval gates to no oversight at all (Table~\ref{tab:hitl-results}). The clearest consequence is at the permissive end: 62.8\% of \mcpapps{} require no approval before a tool executes, the LLM's
request alone is enough to run an enabled tool.The risk is greatest in the 20.0\% of \mcpapps{} with no active gate of any
kind. These applications have neither an allowed list restricting which
tools are enabled nor an approval step, so the LLM can invoke any tool
unconditionally. Within this tier, the 16.4\% that rely on logging alone
retain observability, but no means of prevention, and the remaining 3.5\%
have no HITL mechanism at all. In these configurations, a prompt-injection
or tool-poisoning attack~\cite{Hou25,Hua26}, a compromised MCP
server~\cite{hasan2025model}, or an unexpected LLM decision could trigger
tool execution with no structural opportunity for the user to intervene. This contrasts with how traditional dependency risks are managed. There, invocation is fixed in code and statically analyzable: call sites are auditable, and dependencies can be screened ahead of execution through security scanning and known-vulnerability databases. Dependencies can still misbehave, but the ecosystem provides mechanisms to
surface that risk (e.g., vulnerability scanners and security advisories)
and to contain it (e.g., pinning or removing a flagged dependency) before
any code runs. MCP lacks these safeguards: there is no standard configuration file to scan and no vulnerability database for MCP
servers. Moreover, since the LLM decides at runtime which tools to call,
the only place left to intervene is the tool call itself, yet 62.8\% of
\mcpapps{} have no approval gate there (Table~\ref{tab:hitl-results}).

\subsection{Configuration Patterns Beyond \taxonomy}
 
During manual resolution of \textit{Unsure} configuration labels, we found a few repositories that fit none of the three \taxonomy configuration categories and were absent from our formative sample; we note them as observations. Two patterns recur: \emph{hardcoded} server definitions, where the command or URL is embedded in source code (e.g., \repo{kirillsaidov/ollama-mcp-example}~\cite{ollamamcpexample}), typically in tutorials and demos; and \emph{CLI-supplied} identity, where the server is passed as a command-line argument with no persistent configuration (e.g., \repo{nccgroup/http-mcp-bridge}~\cite{httpmcpbridge}). Both are too rare to affect our findings, but they represent the least visible form of dependency, existing only in runtime invocation and leaving no trace in configuration files or a database.

\section{Threats to Validity}

\paragraph{Construct Validity} This concerns whether \taxonomy{} and our
labeling procedure actually measure what we claim to measure: the MCP
integration practices of \mcpapps. We derived \taxonomy{} through iterative
manual analysis of 50 randomly sampled \mcpapps{} and observed empirical
saturation after approximately 30 repositories; nevertheless, rare
integration strategies may fall outside its categories. Our pipeline is
designed to handle such cases safely: rather than forcing an
out-of-taxonomy repository into an existing category, it returns
\textit{Unsure}, and manual inspection of these cases surfaced only a
handful of rare patterns (e.g., hard-coded server definitions), which we
report in Section~\ref{sec:discussion} and which do not affect our
findings. A second threat is that we measure integration practices through
static source code analysis rather than runtime behavior, so dynamic
configurations may not be detectable from repository contents alone. We
mitigate this by assigning a label only when explicit code evidence
supports it, and conservatively labeling repositories \textit{Unsure}
otherwise.

\paragraph{Internal Validity} This concerns whether our repository mining and classification process introduces any systematic errors. Our study relies on gpt-5 to identify \mcpapps and gpt-5-mini to classify them according to \taxonomy. Recent work on LLM-assisted software engineering highlights several threats associated with such workflows, including prompt sensitivity, hidden model biases, and limited contextual reasoning, especially when tasks require interpretation beyond well-defined coding schemes \cite{Baltes2025,Ernst2026}. These studies also conclude that LLMs are most reliable for deductive classification tasks that employ predefined codebooks and retain human oversight throughout the analysis \cite{Ernst2026}. Our pipeline is designed to mitigate these threats. We designed our pipeline
accordingly. We manually developed \taxonomy{} before any large-scale
classification, giving the model a fixed codebook. We evaluated both the
repository identification pipeline and the taxonomy classification
pipeline against manually annotated ground-truth sets, observing high
agreement with human annotations. When evidence was insufficient, the
model returned \textit{Unsure} rather than a forced prediction, and we
resolved these cases manually. The LLM thus served as an annotation
assistant within a human-supervised workflow.
\paragraph{External Validity} Our study analyzes 1,723 open-source MCP-enabled AI applications hosted on GitHub. Consequently, the observed design patterns may not generalize to proprietary, commercial, or internally developed MCP applications. Furthermore, the MCP ecosystem is evolving rapidly, with new SDKs, transport mechanisms, security practices, and deployment architectures continuing to emerge as the protocol matures. Our findings therefore represent a snapshot of the ecosystem at the time of data collection rather than a characterization of all future MCP-enabled applications.

\section{Related Work}

\subsection{MCP Datasets}

Several authors curated datasets of MCP-related implementations to enable empirical analysis. Toeppe et al.~\cite{MSRdataset} released a dataset of 2,297 GitHub repositories with MCP-relevant implementations, classifying each as server, client, both, or gateway. As discussed in Section~\ref{sec:dataset}, we examined this dataset as a reference point and used it to derive targeted search queries, which were used in our classification pipeline with a focus exclusively on \mcpapps.
Lin et al.~\cite{Lin25} constructed \textit{MCPCorpus}, a large-scale dataset of the MCP ecosystem comprising roughly 14{,}000 MCP servers and 300 clients. For each entry, they record protocol-level attributes (tools, SSE URLs, server launch commands, and related configuration) alongside GitHub repository signals such as star counts. They used a public registry\footnote{\url{https://mcp.so/}} as the primary source for discovering servers and clients, with GitHub used only to enrich the resulting metadata. Two characteristics distinguish their work from ours. First, \textit{MCPCorpus} is predominantly server-centric, only 300 of its entries are clients, whereas ours focuses on the host/\mcpapp{} side. Second, we search GitHub directly for implementations rather than drawing from registries, letting us collect the open-source code of a substantially larger, host-centric dataset. Guo et al.~\cite{Guo25b} also collected a dataset of 8,060 MCP servers and 341 clients as part of a broader measurement study.

\subsection{Empirical Studies and Analysis of MCP}
The majority of MCP research takes a security perspective, focusing on vulnerabilities in MCP servers and the susceptibility of MCP clients to prompt injection and tool-poisoning attacks. Hasan et al.~\cite{hasan2025model} conducted the first large-scale empirical study of 1,899 MCP servers, examining their health, sustainability, security vulnerabilities, and maintainability issues using static analysis and a hybrid analysis pipeline. Hou et al.~\cite{Hou25} conducted a systematic study of MCP's architecture and security landscape, categorizing threats across the server lifecycle
and identifying prompt injection and unauthorized access as primary risks. While HITL mechanisms can be viewed as one defense against unwanted tool calls, we do not specifically examine the security of MCP servers or attack vectors such as prompt injection; instead, we focus on the practices that developers employ at the application layer in \mcpapps.

Guo et al.~\cite{Guo25b} collected 8,060 MCP servers and 341 MCP
clients as part of a measurement study.
As part of their client analysis, they examined the communication
protocols used by clients and whether they support single-server or
multi-server integration.
They found that stdio remains the dominant transport mechanism while
streamable HTTP shows limited adoption, and that 80.9\% of clients
connect to a single server.
While Guo et al.\ characterize the ecosystem at the protocol and topology
level, our analysis extends this by examining how \mcpapps configure their
server lists, how they instantiate MCP client connections, and what
human oversight mechanisms, if any, are in place before a tool is
executed.

Stein~\cite{Stein26} analyzed 177,436 tools published in public MCP server repositories between November 2024 and February 2026, finding a notable shift toward action tools that directly modify external environments (from 27\% to 65\% of tool usage over the study period). This shift to high-stakes action tools shows why the presence or absence of approval gates in \mcpapps has real world consequences, a dimension we study in this paper.

Singh et al.~\cite{Singh25} surveyed MCP's foundational architecture and core primitives, including tools, resources, and prompts. They described tools as capabilities that models can actively invoke at runtime, typically subjected to human approval, distinguishing them from the more passive, predefined behavior of resources and prompts. While Singh et al.\ acknowledged human-in-the-loop (HITL) control as a desirable property of MCP-enabled systems, they do not empirically examine whether and how \mcpapps actually implement such controls in practice. Our work fills this gap through a large-scale empirical analysis, characterizing the diversity of HITL mechanisms that developers employ in practice.

Huang et al.~\cite{Hua26} applied STRIDE and DREAD~\cite{shostack2014threat} frameworks to identify and prioritize potential threats across MCP architectures. They then used their derived threat model to assess 7 MCP-enabled AI applications (hosts), which the authors refer to as MCP clients (Claude desktop for Windows, Cursor, Cline, Continue, Gemini CLI, Claude Code, and Langflow). By creating malicious servers and designing four attack types, they explored which clients are susceptible to the attacks. They specifically looked at different defense or detection mechanisms implemented in the clients: warning messages displayed to the user, confirmation dialogs required, tool execution blocked or sandboxed, or logging of suspicious activity. 
Whereas Huang et al.\ derived defense categories from a threat model and applied them to a small, curated set of hosts, we identify human-in-the-loop mechanisms from a large corpus of \mcpapps{}; our taxonomy partially overlaps with their catalog, both cover blocking approval and logging, and we add allowed-list controls as a third, empirically observed mechanism.

\section{Conclusion}
The Model Context Protocol (MCP) has rapidly become a standard for connecting AI applications with external tools. Yet, how developers integrate MCP servers in practice remains under-studied. We present the first large-scale empirical study of \mcpapps{}, introducing \taxonomy{}, which characterizes MCP integration across configuration, client communication, and human-in-the-loop controls. Applying it to 1{,}723 open-source \mcpapps{} on GitHub via an LLM-assisted mining pipeline, we provide a quantitative characterization of the ecosystem's configuration, communication, and oversight practices. The taxonomy offers a standardized vocabulary for describing MCP integrations, and the released \dataset{} and pipeline lay a foundation for future study of MCP adoption, security, and engineering practices.

\bibliographystyle{IEEETran}
\bibliography{paper}

\end{document}